\documentclass[aps,prl,twocolumn,superscriptaddress]{revtex4}
\usepackage[colorlinks=true,urlcolor=blue,citecolor=blue,linkcolor=blue]{hyperref}
\usepackage{graphicx}
\usepackage{mathrsfs}
\usepackage{bm}
\usepackage{verbatim}
\usepackage{hyperref}
\usepackage{ulem}

\begin{document}
\title{Ideal flat and resolved SU(3) Landau levels in three dimensions}

\author{Mian Peng} \thanks{M.P. and Q.W. contributed equally to this work}
\affiliation{State Key Laboratory of Quantum Optics and Quantum Optics Devices, Institute of Laser spectroscopy, Shanxi University, Taiyuan 030006, China}

\author{Qiang Wei} \thanks{M.P. and Q.W. contributed equally to this work}
\affiliation{Key Laboratory of Materials Physics of Ministry of Education, School of Physics, Zhengzhou University, Zhengzhou, China}
\affiliation{Laboratory of Zhongyuan Light, School of Physics, Zhengzhou University, Zhengzhou 450001, China}

\author{Jiale Yuan}
\affiliation{State Key Laboratory for Extreme Photonics and Instrumentation, College of Optical Science and Engineering, Zhejiang University, Hangzhou, China}
\affiliation{Zhejiang Province Key Laboratory of Quantum Technology and Device, School of Physics, Zhejiang University, Hangzhou, China}

\author{Da-Wei Wang}
\affiliation{Zhejiang Province Key Laboratory of Quantum Technology and Device, School of Physics, Zhejiang University, Hangzhou, China}

\author{Mou Yan}  \thanks{yanmou@zzu.edu.cn} 
\affiliation{Key Laboratory of Materials Physics of Ministry of Education, School of Physics, Zhengzhou University, Zhengzhou, China}
\affiliation{Laboratory of Zhongyuan Light, School of Physics, Zhengzhou University, Zhengzhou 450001, China}

\author{Han Cai} \thanks{hancai@zju.edu.cn} 
\affiliation{State Key Laboratory for Extreme Photonics and Instrumentation, College of Optical Science and Engineering, Zhejiang University, Hangzhou, China}

\author{Gang Chen}  \thanks{chengang971@163.com}
\affiliation{State Key Laboratory of Quantum Optics and Quantum Optics Devices, Institute of Laser spectroscopy, Shanxi University, Taiyuan 030006, China}
\affiliation{Key Laboratory of Materials Physics of Ministry of Education, School of Physics, Zhengzhou University, Zhengzhou, China}
\affiliation{Laboratory of Zhongyuan Light, School of Physics, Zhengzhou University, Zhengzhou 450001, China}

\begin{abstract}
Landau levels (LLs) are of great importance for understanding the quantum Hall effect and associated many-body physics. Recently, their three-dimensional (3D) counterparts, i.e., dispersionless 3D LLs with well-defined quantum numbers, have attracted significant attention but have not yet been reported. Here we theoretically propose and experimentally observe 3D LLs with a sharply quantized spectrum in a diamond acoustic lattice, where the eigenstates are characterized by SU(3) quantum numbers. The engineered inhomogeneous hopping strengths not only introduce pseudomagnetic fields that quantize the nodal lines into LLs but also provide three bosonic degrees of freedom, embedding a generic SU(3) symmetry into the LLs. Using a phased array of acoustic sources, we selectively excite distinct eigenstates within the degenerate LL multiplets and visualize their 3D eigenmodes. Importantly, our approach enables the precise reconstruction of SU(3) quantum numbers directly from eigenmode correlations. Our results establish SU(3) LLs as a tractable model in artificial platforms, and pave the way for synthesizing LLs with zero dispersion and countable quantum numbers in arbitrary dimensions.
\end{abstract}

\maketitle

Landau levels (LLs) of two-dimensional (2D) free electron gases in a magnetic field have been an important topic in condensed matter physics, owing to the topological transport properties and macroscopic degeneracy \cite{Klitzing1980,Klitzing2020}. The flat LLs provide a tractable model for exploring many-body physics \cite{Superconductivity,Coissard2022,Jiaqinature} due to the analyticity of their wavefunctions, which can be resolved by good quantum numbers in certain gauges \cite{Schine_nature,wenxinhua_NP,yanmou_PRL,GuanxiwenRMP,Felser2021}. For example, the single-particle LLs carrying well-defined angular momenta  can be used to construct the Laughlin states \cite{Laughlin1983,Clark2020,Wang2024FQHstate}, which provides an intuitive picture of the fractional quantum Hall effect \cite{Tsui1982,Stormer_RMP}.

Despite these insights into 2D systems, there is an enduring debate on the existence of discrete LLs in three dimensions (3D) \cite{Halperin1987,Koshino2001,Bernevig2007}. The major concern is that the inevitable group velocity along the direction of magnetic field can smear the quantized energy spectrum. Recently, some pioneering works have proposed new mechanisms \cite{Rachel2016PRL,Luhaizhou 3DQHE  Weyl,Jianghua 3DQHE Weyl,3DQHE_CDW} to observe the Hall transport in 3D materials \cite{Xiufaxian 3DQHE,Zhangliyuan 3DQHE,Cheng2024,Weiqiang 3DQHE}. However, further exploration of 3D LLs is hindered by the challenge of achieving a higher-order symmetry beyond the standard SU(2) one \cite{Zhang2001,Li2012,Li2013}. Resolving LLs with quantum numbers associated with this higher-order symmetry is crucial for understanding their interplay with interparticle interactions, potentially enabling exotic phenomena such as 3D fractional quasiparticles and edge-correlated phases \cite{Behnia_Science,Levin2009,Shao2024}.

\begin{figure}[hb!]
\centering
\includegraphics[width=0.8\columnwidth]{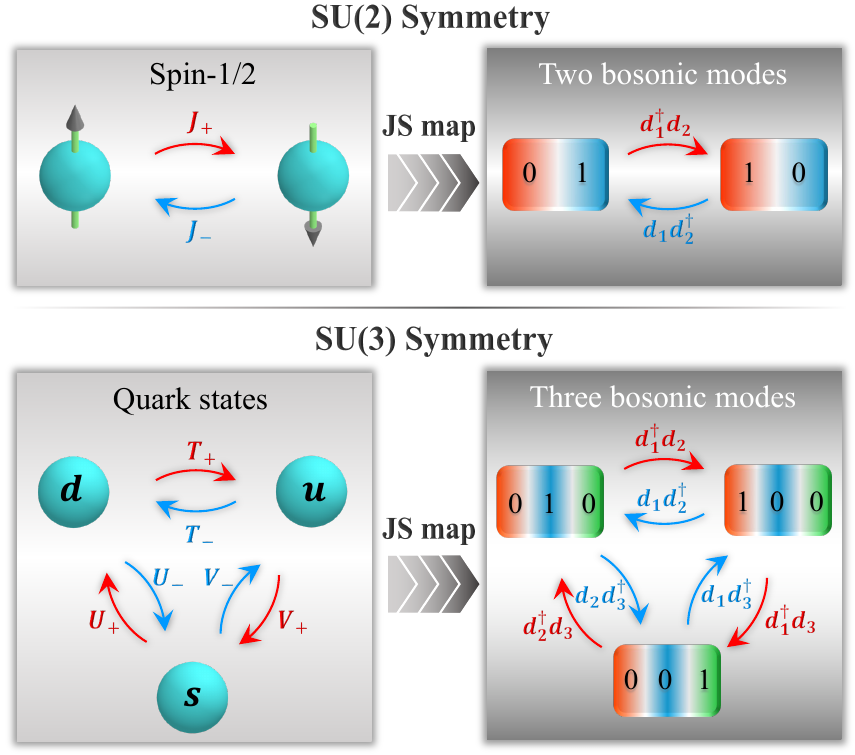}
\caption{Jordan-Schwinger Map. Upper panel: a spin-1/2 can be mapped to two modes sharing one boson. The operator $J_{+}$ ($J_-$)  raises (lowers) the quantum number $J_z=d^{\dagger}_1d_1-d^{\dagger}_2d_2$, equivalent to  swapping the boson between two modes. Lower pannel: the quark states with SU(3) flavor symmetry can be mapped to three modes sharing one boson. The transitions between $u$, $d$, and $s$ quark states shift the two SU($3$) quantum numbers  $T_3=d^{\dagger}_1d_1-d^{\dagger}_3d_3$ and $Y=d^{\dagger}_1d_1-2d^{\dagger}_2d_2+d^{\dagger}_3d_3$.
}
\label{Fig1}
\end{figure}

\begin{figure*}[ht]
\centering
\includegraphics[width=14cm]{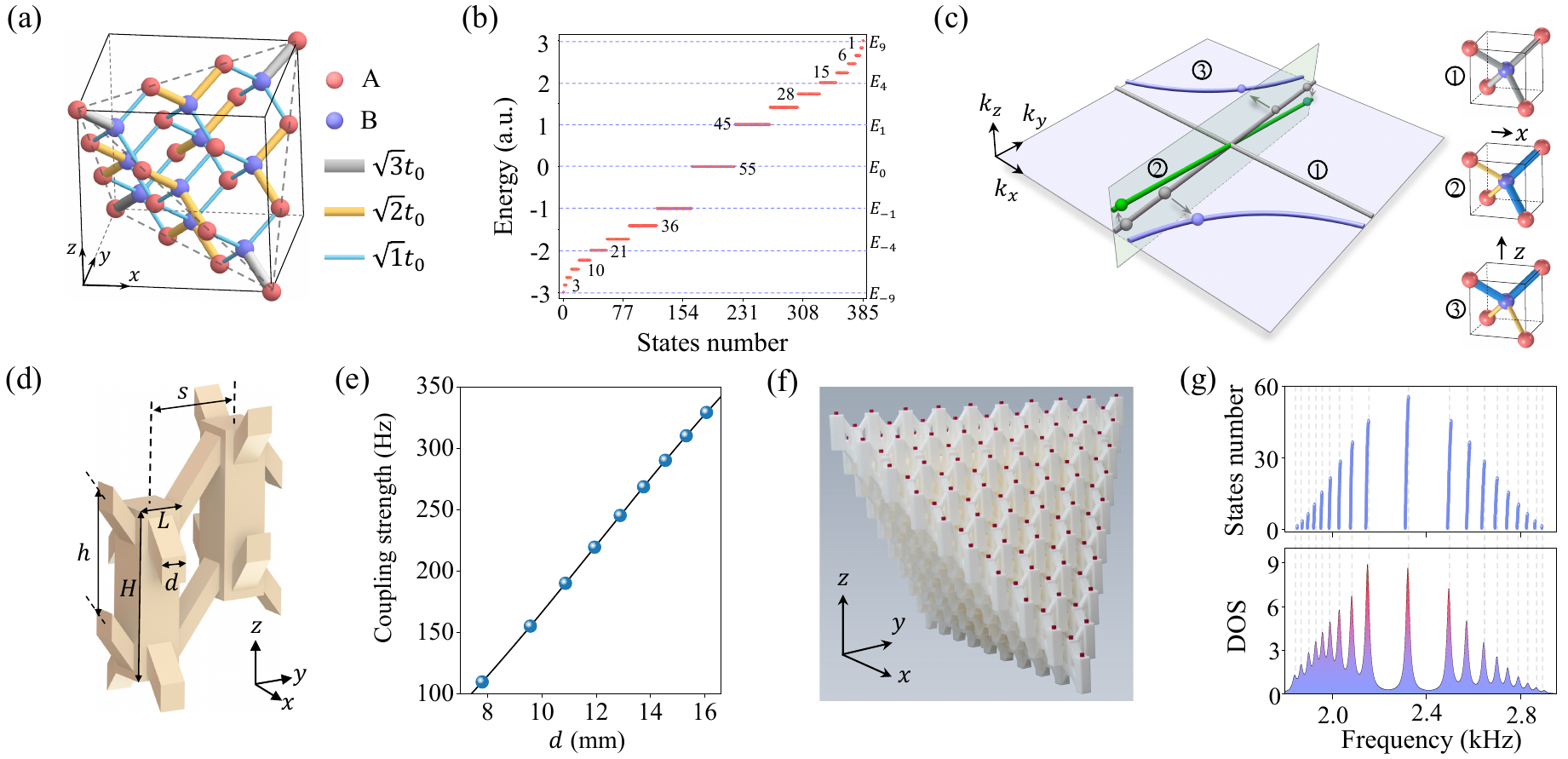}
\caption{(a) Schematics of the engineered lattice ($N=3$) with site-dependent hopping strength. (b) The energy spectrum for $N=9$. The inserted numbers denote the degeneracy of each LL. (c) The nodal lines when the unit cells are marked in 1 (gray lines), 2 (green lines) and 3 (purple lines). The shift of the Dirac nodal point (spheres) are denoted by arrows. (d) Acoustic two-cavity system as a unit cell. (e) Simulated parameter fitting of the acoustic coupling strength with respect to the tube side length $d$. (f) A photo of the 3D-printed acoustic lattice with $N=9$. (g) Density of states (DOS) of the LLs. Upper panel: the simulated degenerate numbers for the sample in (f). Lower panel: the measured DOS spectrum as a function of the excitation frequency.
}	
\label{Fig2}
\end{figure*}

In this letter, we report the realization of completely flat and quantum-number-resolved LLs in a 3D diamond acoustic lattice with inhomogeneous square-root hoppings. The carefully designed inhomogeneity embeds a desired SU(3) symmetry. It not only introduces pseudomagnetic fields (PMFs) \cite{Guinea2010,Levy2010,Gomes2012,Rechtsman2013,Yang2024,Barczyk2024,Barsukova2024} by quantizing the Dirac nodal lines, but also provides three bosonic degrees of freedom to each LL. Three redundant bosonic modes, as highlighted in Fig.~\ref{Fig1}, construct a generic SU(3) symmetry via the Jordan-Schwinger map. After characterizing the degenerate multiplets with two SU(3) quantum numbers, we can selectively excite distinct localized eigenstates and visualize their 3D eigenmodes. We also directly obtain two quantum numbers by inspecting the eigenmode correlation, which describes the relative phase between neighboring sites. This method is similar to characterizing the angular momentum of 2D LLs by counting the winding number of their chiral phase correlation \cite{Schine_nature,Clark2020,Schine2019}. Our theoretical simulation, and experimental results are in good agreement, and can be extended to the systems in arbitrary dimensions with SU($M\geq3$) symmetry.

Figure~\ref{Fig2}(a)  schematically shows a 3D diamond lattice shaped as a tetrahedron with $N=3$ ($N$ determines the size of the lattice). The two sublattices (labeled A and B) are connected by four bond vectors $\bm{\delta}_1=(1,-1,-1)$, $\bm{\delta}_2=(-1,1,-1)$, $\bm{\delta}_3=(-1,-1,1)$, and $\bm{\delta}_4=(1,1,1)$. The position of each lattice site can be denoted by four non-negative integers as $\bm{r}=\sum_{j=1}^4 n_j\bm{\delta}_j$ ($n_j>0$). The B-sublattice site at position $\bm{r}$ is coupled to the four nearest-neighbor A-sublattice sites at position $\bm{r}+\bm{\delta}_j$ with hopping strength $t_j(\bm{r}) = t_0\sqrt{n_j+1}$.  The lattice boundary is formed by the constraint $\sum_{j=1}^4 n_j=N-n_s$ with $n_s=0$ $(1)$ for the A (B)-sublattice.
The tight-binding Hamiltonian is written as
\begin{equation}
H_{\text{TB}} = \sum_{\bm{r},j}t_j(\bm{r}) B_{\bm{r}}^\dagger A_{\bm{r}+\bm{\delta}_j} + \text{H.c.},  \label{hamil}
\end{equation}  
where H.c. represents the Hermitian conjugate.

We obtain the square-root-scaling discrete spectrum in Fig.~\ref{Fig2}(b) by diagonalizing $H_{\text{TB}}$. Such an ideal quantization is induced by the inhomogeneity of $t_j(\bm{r})$, which can be understood in the continuum limit near the tetrahedron center. The $H_{\text{TB}}$ is locally isotropic at $\bm{r}=0$, exhibiting intersecting Dirac nodal lines \cite{Burkov2011,Heikkila2011,Takahashi2013}. By linearizing  the hopping strength variation as $\nabla_{\bm{r}} t_j(\bm{r})\propto \bm{\delta}_j$, we can obtain the nodal lines in momentum space as a function of $\bm{r}$ locally, which is exemplified by the local unit cells along $x$ and $z$ axes in Fig.~\ref{Fig2}(c). As a result, each Dirac point along the $i$-axis ($i=x,y,z$) is then effectively subject to minimal coupling with a pseudovector potential, whose curl yields an $i$-directional PMF. Since both the strength and direction of the PMF vary with respect to the three nodal lines \cite{Cai2023,Kohler2024}, such LLs cannot be simply reduced to a stack of 2D systems where only out-of-plane magnetic fields matter.


The observation of PMF induced LLs is implemented in an acoustic crystal with a lattice constant of $a=84.86$ mm. The unit cell [Fig.~\ref{Fig2}(d)] consists of two square acoustic cavities (side length $L=20$ mm and height $H=75$ mm) and four pairs of narrow connecting  tubes with varying side lengths $d$. The projected distance of the two cavities along the $x$, $y$ and $z$ direction is $s/\sqrt{2}=30$ mm, and the vertical distance between the two narrow tubes is $h= (0.9H-d)$ mm. The desired site-dependent hopping strength can be achieved by varying the side length $d$ of each pair of tubes. In Fig.~\ref{Fig2}(e), we simulate the acoustic coupling strength versus the side length d of the coupling tubes (black curve), satisfying $t_A=(-0.06d^3+2.01d^2+3.21d-11.6)$ Hz. According to the relation,we present nine different sizes of $d$ (blue circles) to achieve the required square-root-scaling acoustic couplings (see details in  Supplementary Materials (SM) \cite{SM}).

Assembling these unit cells, we 3D-print a tetrahedral sample [Fig.~\ref{Fig2}(f)] with a total length of $763.68$ mm, containing $385$ cavities, which corresponds to  equation (\ref{hamil}) with $N=9$.
We simulate the eigen-spectrum [upper panel in Fig.~\ref{Fig2}(g)] and notice that the LLs with positive indices show a slightly larger bandgap than those with negative indices. This asymmetry can be attributed to the inevitable next-nearest-neighbor coupling \cite{SM}. To observe the discrete spectrum, we measure the density of states (DOS) by probing the acoustic response spectrum of each cavity. In experiments, we place the acoustic point source and detector in the same cavity simultaneously. We then scan the acoustic frequency from $1.4$ kHz to $3.4$ kHz and collect the response signal in a network analyzer. Lower panel in Fig.~\ref{Fig2}(g) shows the measured DOS spectrum as a function of frequency. The peak positions, associated to the frequency of each LL, correspond well with the simulated spectrum.

\begin{figure}[ht]
\centering
\includegraphics[width=0.9\columnwidth]{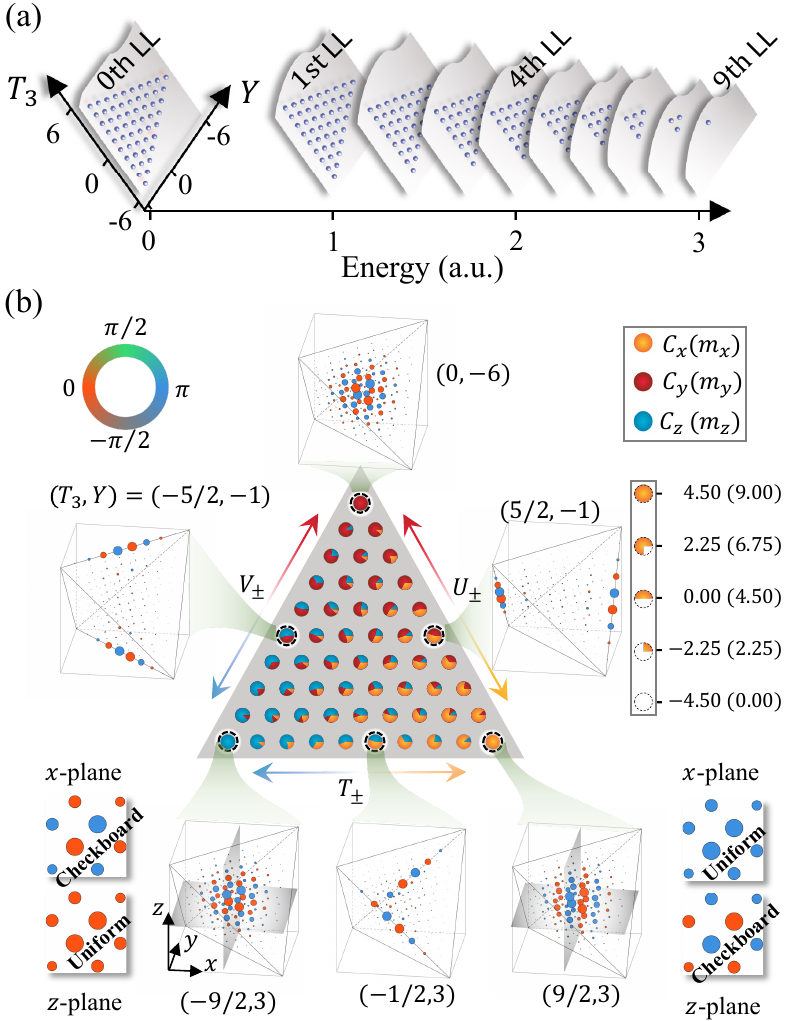}
\caption{ (a) The LL multiplets as a function of the energy. (b) The multiplet, eigenmode correlation, and SU(3) quantum numbers for the 0th LL. For each eigenmode, the values of three in-plane correlation $C_j$ ($j=x,y,z$) are scaled as the size of colored segments, respectively. Several typical eigenmodes are plotted on the side to illustrate the field distribution. For the 0th LL, the  eigenmode at the three vertices of the multiplets are similar in the strength distribution, only rotate their phase pattern (insets) from uniform to checkerboard in $x$- and $z$-planes marked by the shadow.
}    \label{fig4}
\end{figure}

\begin{figure*}
\centering
\includegraphics[width=16cm]{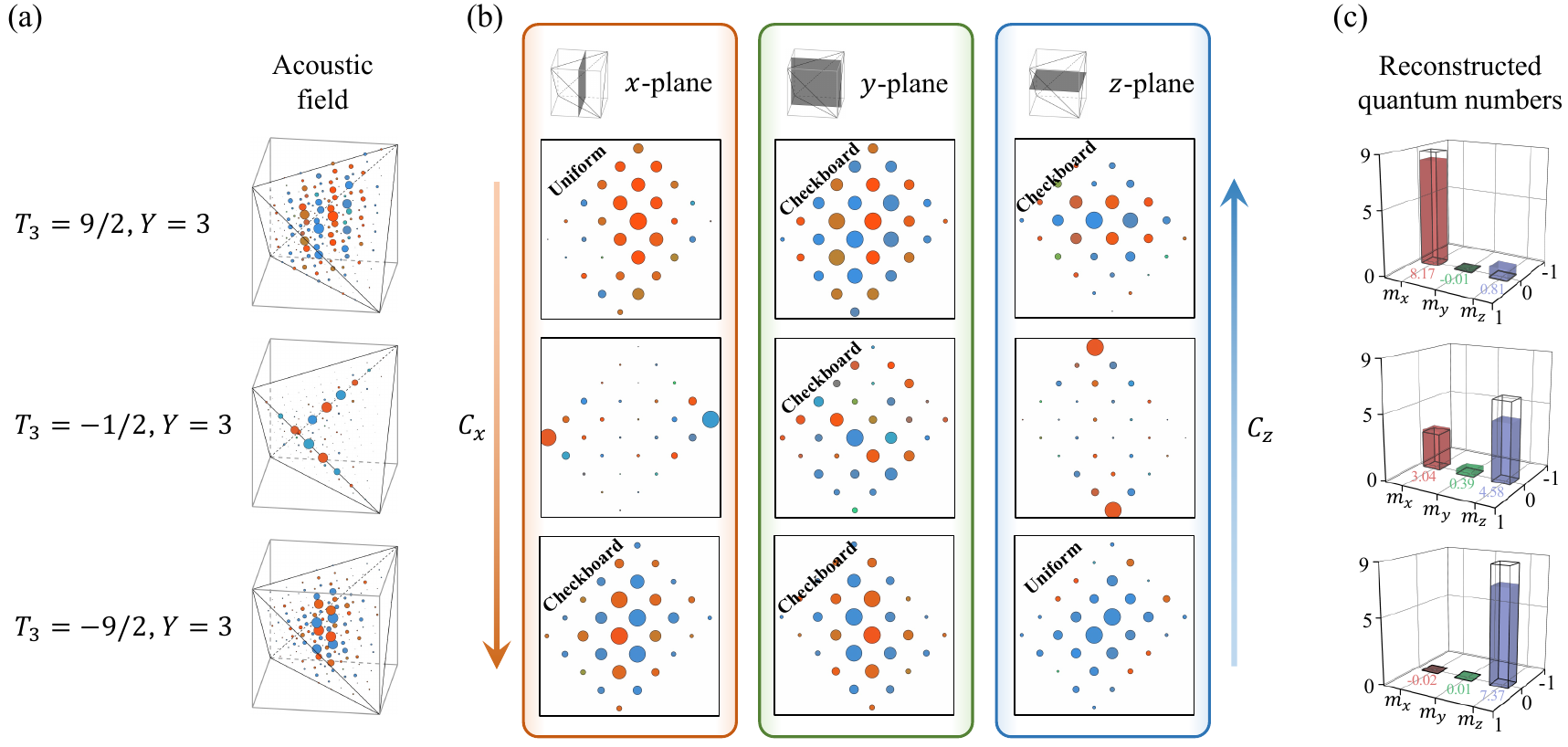}
\caption{(a) Measured excited acoustic pressure field distribution. The color and size of dots represent the phase and amplitude, respectively. (b) The corresponding in-plane phase pattern in the $x$-, $y$- and $z$-planes marked by shadow. The size of dots are normalized with respect to the maximum amplitude in the plane. The arrows shows the descending direction of $C_x$ and $C_z$, switching the phase pattern from uniform to checkerboard.   (c) The ancillary quantum numbers reconstructed from the correlation of the excited field distribution. The empty box denotes the simulated results, while the filled bar represents the experimental data. $(T_3,Y) = (9/2,3)$, $(-1/2,3)$, and $(-9/2,3)$ from uppper to lower panels.}    \label{fig5}
\end{figure*}

However, such a perturbative approach  commonly taken in strain engineering is not sufficient to investigate the hidden symmetry of LLs. Instead, we need to consider the inhomogeneity from a global viewpoint, i.e.,  mapping the tight-binding Hamiltonian $H_{\text{TB}}$ to an exactly solved quantum spin-boson model \cite{Cai2021,Deng2022}. The A(B)-sublattice site at position $\bm{r} (n_1,n_2,n_3,n_4)$ can be mapped into a configuration by assigning $N$ particles into four bosonic modes and a spin, denoted as $\left|\downarrow(\uparrow), n_1,n_2,n_3,n_4\right\rangle$. Now the constraint $\sum_jn_j=N-n_s$ represents particle number conservation, with $n_s=1(0)$ denoting the spin-up (down) state and $n_j$ being the particle number in the $c_j$ mode, respectively. Therefore, the hopping between sites along $\bm{\delta}_j$ can be interpreted to  annihilating or creating a particle from $c_j$ mode and simultaneously flipping the spin. In particular, the factor $\sqrt{n}$ attached to the hopping strength of $H_{\text{TB}}$ originates from the quantum nature of bosons, i.e., the transition rate of putting one more boson into a Fock state increases with its existing occupation number as $c^\dagger|n\rangle=\sqrt{n+1}|n+1\rangle$. Therefore, the properties of the tight-binding lattice $H_{\text{TB}}$ can be fully captured by $H_{\text{Q}}=t_0\sum_j(c_j\left|\uparrow\right\rangle \left\langle \downarrow\right|+\text{H.c.})$, where $c_j$ ($c_j^\dagger$) is the $j$th annihilation (creation) operator \cite{SM}.

The  quantum spin-boson model offers an elegant way to obtain the quantized eigen-spectrum in Fig.~\ref{Fig2}(b) and examine embedded symmetry in Fig.~\ref{fig4}. First, we introduce a collective bright mode $d_0=1/2 \sum_j c_j$ to simplify $ H_{\text{Q}} $ to a single-mode Jaynes-Cummings model \cite{Perez2010,Keil2011}. Depending on the particle number $m$  in the bright mode, the energy spectrum is shapely quantized as $E_{\pm{m}}=\pm{2t_0\sqrt{m}}$ ($m=0,1,2\dots N$), where  $\pm m$ is also the LL index. Besides $d_0$, there are three redundant dark modes $d_x=(c_1+c_2-c_3-c_4)/2$, $d_y=(c_1-c_2+c_3-c_4 )/2$, and $d_z= (c_1-c_2-c_3+c_4)/2$ with $[d_{x,y,z},H_{\text{Q}}]=0$. They provide three-fold degrees of freedom to impose the SU($3$) symmetry on the LLs, which is evidenced by the degeneracy of LLs. For the $m$th LL, there are $(N-m+1)(N-m+2)/2$ possible states by assigning ($N-m$) particles into three dark modes  $(m_x,m_y,m_z)$. The Jordan-Schwinger map allows us to label these degenerate states by two quantum numbers, termed “isospin” $T_3=(m_x-m_z)/2$ and “hypercharge”  $Y=(m_x-2m_y+m_z)/3$,  analogous to the flavor symmetry of quark states \cite{Georgi}.  

To highlight the symmetry, we show that the LLs here exhibit all the general features of SU(3) algebra {\cite{Zee}}. In the $T_3$-$Y$ plane, it contains three subalgebras along three lines oriented 120 degrees from each other, each of which is isomorphic to the SU(2) algebra. As known from angular momentum theory, there are rising and lower operators for each subalgebra. The operators $T_+ = T_-^\dagger=d_x^\dagger d_y$ act on the state to form stripes of $T$-multiplets parallel to the $T_3$-axis, counted by the quantum number $T_3\equiv[T_+,T_-]/2$. This process  can be viewed as transferring particles between $d_x$ and $d_y$ modes. Along the other two axes, there are $U$- and $V$-multiplets formed by the associated operators $U_+=U_-^\dagger=d_y^\dagger d_z$ and $V_+=V_-^\dagger=d_x^\dagger d_z$, respectively. The three sets of SU(2) multiplets are coupled because of the commutation relation, such as $[T_+,V_- ]=-U_-$. This coupling leaves only two independent quantum numbers, $T_3$ and $Y\equiv ([U_+,U_-]-[V_+,V_-])/3$. Therefore, they combine to form  a finite SU($3$) multiplet as a regular triangle in the $T_3$-$Y$ plane. 


This SU($3$) symmetry is also manifested in the eigenmode field distribution $\psi_{\bm{r}}$, which can be obtained by reversely transforming the bright-dark-mode Fock states back to the Fock space of $c_j$ modes \cite{SM}. In a 2D scenario, the LLs carry quantized angular momentum, counted by the chiral winding number of the phase correlation \cite{Schine2019}. Inspired by that, we demonstrate that the phase correlation of 3D LLs can also be used to extract the ancillary numbers $(m_x,m_y,m_z)$ and also the SU($3$) quantum numbers $(T_3,Y)$ for the $m$th LL  \cite{SM}
\begin{equation}
m_i=C_i+N/2-m,  \label{quantum_number}
\end{equation}
where $C_i$ is the summation of all the correlation terms spanned the $i$-planes, i.e., $C_x=\sum\limits_{\bm{r}}(C_{\bm{r},\bm{\delta}_{14}}+C_{\bm{r},\bm{\delta}_{23}})$, $C_y=\sum\limits_{\bm{r}}(C_{\bm{r},\bm{\delta}_{13}}+C_{\bm{r},\bm{\delta}_{24}})$, and $C_z=\sum\limits_{\bm{r}}(C_{\bm{r},\bm{\delta}_{12}}+C_{\bm{r},\bm{\delta}_{34}})$. Here $C_{\bm{r},\bm{\delta}_{jj'}} = \sqrt{n_{j'}(n_j+1)}\text{Re}\left[\psi^*_{\bm{r}}\psi_{\bm{r}+\bm{\delta}_{jj'}} \right]$ denotes the correlation between two sites with $\bm{\delta}_{jj'}= \bm{\delta}_j-\bm{\delta}_{j'}$.

Equation {\color{blue}(\ref{quantum_number})} can be illustrated by the 0th LL multiplet in Fig.~\ref{fig4}(b). We consider the eigenmode at the right vertex with $(T_3,Y)=(9/2,3)$. The field phase $\text{Arg}\left[\psi_{\bm{r}}\right]$ takes on values of either 0 or $\pi$. In $x$-planes, the phase is uniform, analogous to ferromagnetism, holds the maximum in-plane correlation, which coincides with the maximized $m_x=9$. In $y$- and $z$-planes, the checkerboard phase pattern, analogous to antiferromagnetism, exhibits the minimum correlation, corresponding to $m_y=m_z=0$.  Along the $T_3$-axis, the checkerboard pattern in $y$-planes preserves to keep $m_y$ unchanged, while $C_x$ and $C_z$  swap to flip $m_x$ and $m_z$. Additionally, we also illustrate the 8th LL in SM \cite{SM}, forming a smallest non-trivial representation of SU($3$) with only three states.


In experiments, we resolve different eigenmodes by selectively exciting the acoustic pressure field. Since their amplitude distribution is localized [Fig.~\ref{fig4}(b)], the excitation process can be  implemented by four point sources with engineered relative phases \cite{SM}. We exemplify the selective excitation with the eigenmode on the right vertex of the 0th LL multiplet with $(T_3,Y)=(9/2,3)$. Its amplitude distribution is a Gaussian-like function centered at the origin. We place the point sources at the four sites with the largest amplitudes, match the phase distribution (two with $0$ phase while  other two carry $\pi$ phase), and fix the excitation frequency of $2.32$ kHz around the $0$th LL (more details in SM \cite{SM}). We then measure the acoustic pressure field in each acoustic cavity [upper panel of {Fig.~\ref{fig5}(a)].  We demonstrate the checkerboard pattern in the $y$- and $z$-planes, and a uniform pattern in the $x$-plane [Fig.~\ref{fig5}(b)]. According to the measured acoustic field, we  obtain the correlation $(C_x, C_y, C_z)=(4.32, -4.45, -4.55)$, and the associated ancillary numbers $(m_x, m_y, m_z)=(8.17, -0.01, 0.80)$, as shown by the upper panel in {Fig.~\ref{fig5}(c). Hence, the experimentally reconstructed SU($3$) quantum numbers $(T_3,Y)$ is $(3.68, 3.00)$.

The eigenmodes with $(T_3,Y)=(-1/2,3)$ and $(-9/2,3)$ along the $T_3$-axis are also resolved.  The acoustic pressure field are shown in Figs.~\ref{fig5} as middle and lower panels, respectively. The measured correlation leads to ancillary numbers $(m_x, m_y, m_z)=(3.04, 0.39, 4.58)$ and $(-0.02, 0.01, 7.37)$, yielding the reconstructed $(T_3,Y)$ values are $(-0.77, 2.27)$ and $(-3.69,2.44)$. In addition, we execute the similar process for the $8$th LL, which contains three states to form a smallest non-trivial representation of SU($3$), and the experimental data are shown in SM \cite{SM}.



In conclusion, we have theoretically predicted and experimentally demonstrated the existence of SU(3) LLs, of which  eigenmodes can be selectively probed.  Each eigenmode can be well defined by two quantum numbers, which serve as the signature of the embedded SU(3) symmetry. More importantly, we can determine these quantum numbers by measuring the correlation of different eigenmodes. Our work not only confirms the characteristics of SU(3) algebra but also opens new avenues for exploring higher-dimensional quantum Hall effect in artificial platforms \cite{Taie2022,Tsui2022} and even impacting the quantum simulation of particle physics \cite{Atas2023,Bauer2023}.

\begin{acknowledgments}
This work is supported partly by the National Key R \& D Program of China under Grant (Nos. 2022YFA1404500 and 2021YFA1400900),
Cross-disciplinary innovative research group project of Henan province (No. 232300421004), the National Natural Science Foundation of China (Nos. 12374335, U21A20437, 12074232, 12125406, 12104450, and 12204290), China Postdoctoral Science Foundation (Nos. BX20220195 and 2023M732146), and Zhejiang Provincial Natural Science Foundation of China (No. LZ24A040001).
\end{acknowledgments}

\end{document}